\documentclass[a4paper,notitlepage,english]{article}

\usepackage[utf8]{inputenc}
\usepackage[LGR,T1]{fontenc}        % T1 font encoding en support voor Griekse letters
\usepackage[british]{babel}
\usepackage[iso]{isodate}           % Normale datums

\usepackage[top=2.5cm,bottom=2.5cm,left=2cm,right=2cm]{geometry} % Marges
\usepackage{lastpage}               % Paginanummers in footer
\usepackage[compact]{titlesec}	    % Compactere headings
\usepackage[bottom]{footmisc}       % Voetnoten onderaan pagina zetten

\usepackage[babel=true,final]{microtype}	% Microtypography
\usepackage{parskip}                        % Normale paragrafen
\usepackage{lmodern}
\usepackage{textgreek}
\usepackage{hyphenat}

\usepackage{hyperref}    % Klikbare verwijzingen
\usepackage{color}
\usepackage[dvipsnames]{xcolor}		% Kleurnamen
\usepackage{graphicx}               % Plaatjes
\usepackage{float}                  % Whatever floats your goat
\usepackage{caption}
\usepackage{subcaption}
\usepackage{amsmath}
\usepackage{abstract}
\usepackage{authblk}
\usepackage{enumitem}

\usepackage{tabulary}				% Betere tabellen
\usepackage{booktabs}               % Same
\usepackage{multirow}               % Rowspan

\usepackage[parse-numbers=false]{siunitx}

\usepackage[sorting=none,sortcites,giveninits,maxnames=5,backend=biber,urldate=iso,date=year]{biblatex} % Beter dan BibTex
\usepackage{csquotes}				% Iets met BibLaTeX en Hyperref

\usepackage[a-2b]{pdfx}

\definecolor{CoolBlack}{rgb}{0.0, 0.18, 0.39}
\hypersetup{
	hidelinks,
	colorlinks,
	linkcolor  = CoolBlack,
	citecolor  = CoolBlack,
	urlcolor   = OliveGreen
}

\setlist{leftmargin=1em}

\newcommand{\name}{OpenHumidistat}

\title{\name{}:\\Humidity-controlled experiments for everyone}
\author{Lars B. Veldscholte}
\author{Sissi de Beer}
\affil{Sustainable Polymer Chemistry Group, Department of Molecules \& Materials\\MESA+ Institute for Nanotechnology, University of Twente\\P.O. Box 217, 7500 AE Enschede, the Netherlands}
\date{\today}

\begin{document}
	\maketitle

	\begin{abstract}
		\vspace{1em}
		Humidity control is a crucial element for a wide variety of experiments. Yet, often naive methods are used that do not yield stable regulation of the humidity, are slow, or are inflexible. PID-based electropneumatic humidistats solve these problems, but commercial devices are not widespread, typically proprietary and\slash or prohibitively expensive. Here we describe \name{}: a free and open-source humidistat for laboratory-scale humidity control that is affordable (<€500) and easy to build. The design is based around mixing a humid and dry air flow in varying proportions, using proportional solenoid valves and flow sensors to control flow rates. The mixed flow is led into a measurement chamber, which contains a humidity sensor to provide feedback to the controller, to achieve closed-loop humidity control.
	\end{abstract}

\vspace{1em}
\textbf{Keywords:} humidity, controller, electronics, pneumatics, PID
\vspace{1em}

\begin{figure}[H]
	\centering
	\includegraphics[width=0.8\linewidth]{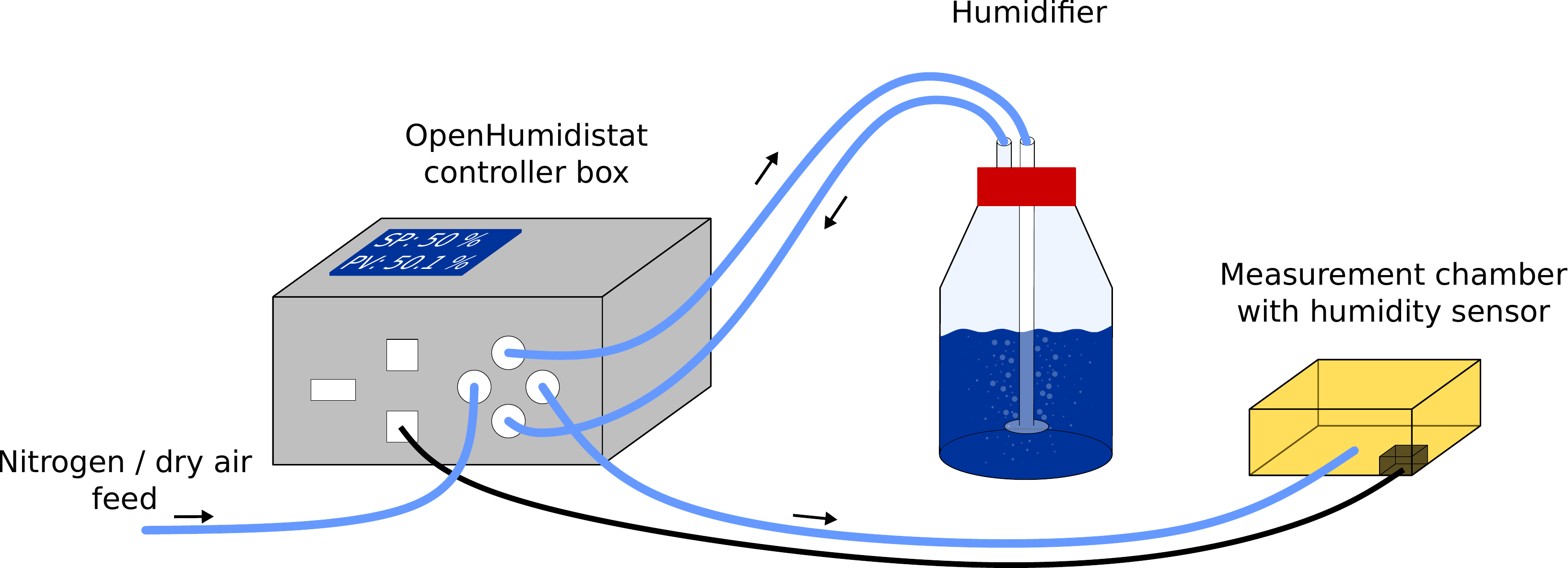}
\end{figure}

\twocolumn
\section{Hardware in context}
% Applications of humidistats
Humidity control is essential for a plethora of biological, chemical, and physical experiments~\cite{johansson_precise_2021}, some of which simply require a stable, constant humidity, while in others the humidity might be an experimental variable which is varied. In either case, careful humidity control is necessary. A device that accomplishes this is called a humidity controller or \emph{humidistat} (analogous to a thermostat for temperature).

% Different humidity control methods
Systems designed to adjust the humidity in experiments are often based on equilibria of water vapour with salt solutions \cite{Greenspan1977, Brio2020}, local heating of a water bath in a reservoir chamber \cite{Perino2002, Micciulla2019}, or using an air stream composed of a mixture of approximately dry and water-saturated air \cite{deBeer2010, Chhabra2013, Karpitschka2017, Atesci2018} that is manually mixed in varying ratios. The first method provides stable control, but is very inflexible: for every value of humidity, a specific salt is required. This moreover means that when an experiment calls for sweeping a range of humidities, the salt solution has to be manually swapped out for another one when moving to the next humidity point, interrupting the experiment~\cite{gaponenko_open_2019}. Depending on the size of the chamber, equilibration also can take quite long. For this reason, this method is mainly suited for experiments which only require long-term stabilisation of the humidity on a single value in a air-tight chamber. The other methods are more suited to experiments requiring changing humidities, but typically yield poor control due to the lack of (automatic) feedback. This, combined with the typically long response times makes them arduous to regulate manually and prone to external disturbances.

Therefore, humidistats based on electropneumatic devices have been developed. In these devices, electronically\hyp{}driven valves are used to mix a humid and dry air stream in varying proportions. A humidity sensor in a chamber is used to provide feedback to the controller actuating the valves. The advantages of such humidistats are fast settling times, good disturbance rejection, broad attainable humidity range, flexibility and portability, and easy, intuitive operation.
% Why DIY
Electropneumatic humidistats are commercially available, but are often unsuitable to many labs because they are typically prohibitively expensive, proprietary, and\slash or tailored to a specific experimental setup.

% Comparison with Boulogne
An open-source design of an electropneumatic humidistat using solenoid valves driven by an Arduino was published by Boulogne~\cite{boulogne_cheap_2019}. This was followed by the HumidOSH~\cite{lau_humidosh_2020}, which is a self-contained environmental chamber utilising a similar principle. In this design, ordinary (non-proportional) solenoid valves are opened\slash closed for varying time periods to control the ratio of humid and dry air led into a chamber. This control method, called time-proportional control, can be regarded as PWM (Pulse-Width Modulation), except on longer timescales. This method then relies on the inherent (diffusion-driven) time averaging properties of the comparatively long residence time of the chamber to attain a stable humidity. For this reason, the minimum settling times are inherently limited: Boulogne reports settling times on the order of \SI{1/2}{\hour}, and the HumidOSH takes several days. Reducing the residence time of the chamber is not an option since in that case, the resulting chamber humidity will fluctuate significantly over time. Hence, when quicker dynamics or small chambers\footnote{E.g. those with a volume of several \si{\centi\liter}, as is usual for cells used for analytical techniques such as contact angle measurement, quartz crystal microbalance, ellipsometry, atomic force microscopy, neutron reflectometry, etc.} are desired, a more delicate control technique is required.

% Comparison with v1
In our previous publication~\cite{veldscholte_design_2021}, we presented an affordable open-source humidistat for lab-scale applications based on an Arduino microcontroller, a set of proportional solenoid valves, a gas washing bottle, and a humidity sensor. Using closed-loop feedback, humidity control is achieved in a small measurement chamber by regulating the ratio of dry and humid air flows. Because this design utilises \emph{proportional} solenoid valves which allow for analog control of the flow rate through the valve, stable humidity control can be achieved even in small chambers. It achieved fast regulation of humidity over a wide range (10–95\%) with settling times within \SI{30}{\second}.

\begin{figure*}[ht]
	\centering
	\includegraphics[width=\linewidth]{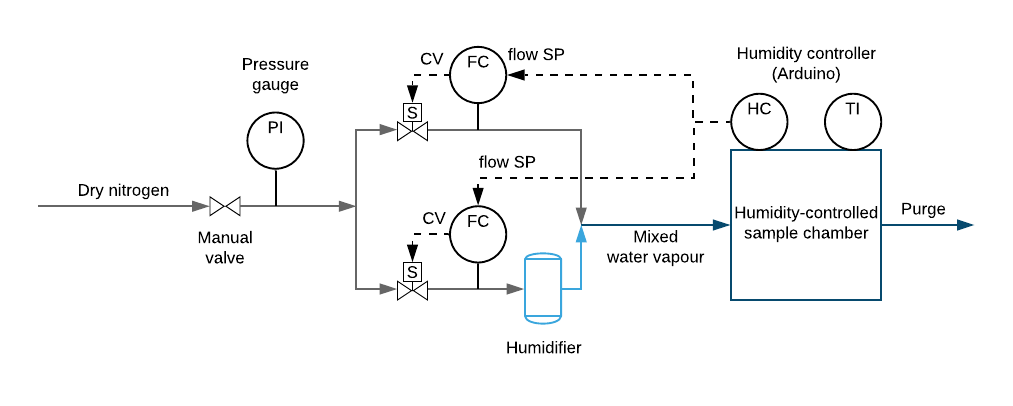}
	\caption{Piping and instrumentation diagram of the humidistat. Starting from the left, pressure-regulated dry nitrogen is fed into the device. The stream is split into two. One of the streams is humidified, while the other is not. Solenoid valves inserted in the streams modulate the flow rates of humidified and dry nitrogen, after which they are combined again and led into the chamber. A controller acting on the humidity in the chamber sets setpoints of flow controllers that determine the ratio of humid and dry flows. PI, TI, FC, and HC stand for pressure indicator, temperature indicator, flow controller, and humidity controller, respectively.}
	\label{fig:piping}
\end{figure*}

% This work
Here, we improve upon that design, primarily by moving to cascade control, in which the individual flow rates are closed-loop controlled as well for improved robustness and dynamic performance. This is comparable to the approach taken by Gaponenko et al.~\cite{gaponenko_open_2019}, but avoiding the use of (expensive) off-the-shelf mass flow controllers. Furthermore, many small improvements to both the hardware and firmware are made, most notably an upgraded humidity sensor, a much larger graphical display, and a more powerful microcontroller. Meanwhile, the firmware is (re)written to include the cascade PID control loop with feed-forward and a new graphical user interface to match the new display, while care is taken to keep the code modular and fully configurable, such that it is compatible with (any combination of) both `original' and upgraded components. Finally, by including comprehensive build instructions, printed circuit board (PCB) designs for the electronic circuits, and 3D printable models for the enclosure, this paper also facilitates reproducibility of the device.

\section{Hardware description}
\subsection{Principle}
The working principle behind the humidistat is the mixing of a humid and dry airflow in varying proportions. In order to adjust the flow rates of humid and dry air, proportional solenoid valves are utilised, which provide analog electronical control over the flow rates. A PID controller running on a Teensy LC microcontroller board~\cite{teensylc} actuates the valves in order to attain the desired humidity in a chamber, which contains a humidity sensor to provide feedback to the controller (\autoref{fig:piping}).

\subsubsection{PID control}
PID control is a closed-loop control scheme, which means that feedback is employed to control some process variable (PV from here on). The letters P, I, and D stand for proportional, integral, and derivative respectively, and refer to the three actions included in the control scheme. The PID controller seeks to close the error between the PV and the user-desired value called the setpoint (SP) by actuating some final control element (FCE) whose state is represented by a control variable (CV). Often, the FCE is a valve, or sometimes a electric device such as a heater. \cite{kuphaldt_lessons_2017, wescott2018, visioli_practical_2006}

\begin{figure*}[ht]
	\centering
	\includegraphics[width=\linewidth]{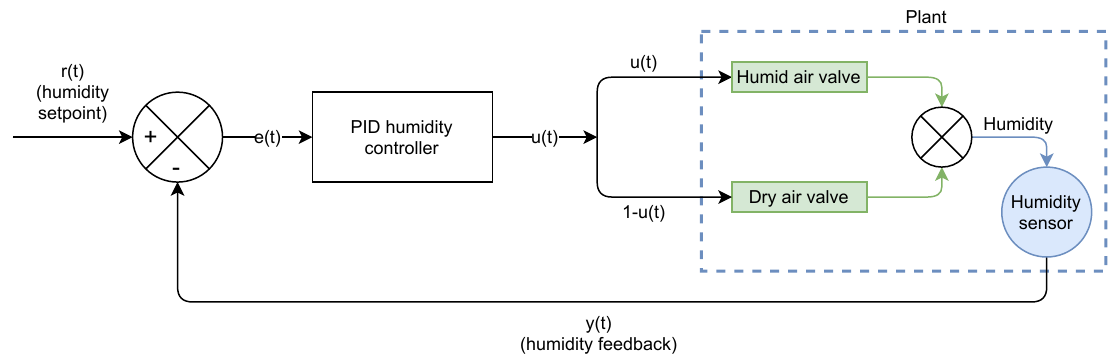}
	\caption{Block diagram of the single control loop.}
	\label{fig:pid_1}
\end{figure*}

\begin{figure*}[ht]
	\centering
	\includegraphics[width=\linewidth]{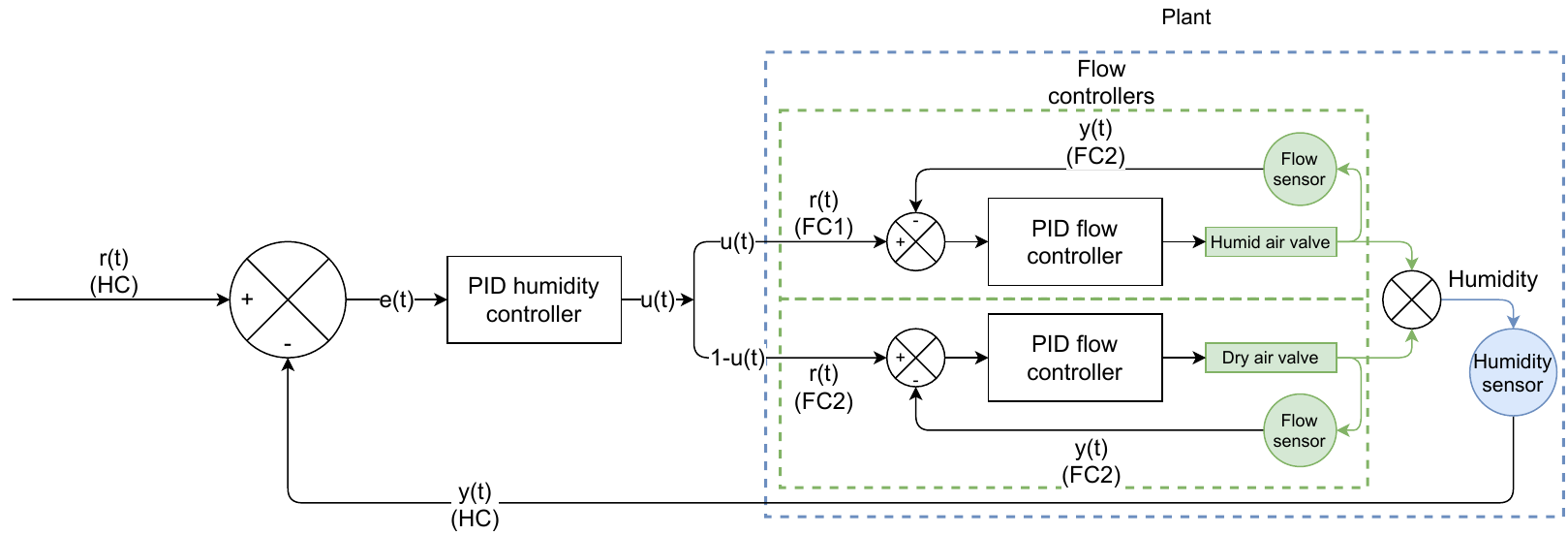}
	\caption{Block diagram of the cascade control loop.}
	\label{fig:pid_2}
\end{figure*}

The CV is the sum of the three principal components of the PID controller: the proportional term, which is given by the instantaneous error between the PV and SP; the integral term, which is given by the integral of the error over time; and the derivative term, which is given by the rate of change of the error with respect to time. Each term is multiplied by its respective gain ($K_{\rm p}$, $K_{\rm i}$, and $K_{\rm d}$), as expressed by the following equation:

\begin{equation}
	u(t) = K_{\rm p}\ e(t) + K_{\rm i} \int_0^t e(x)\ \mathrm{d}x + K_{\rm d} \frac{\mathrm{d}e(t)}{\mathrm{d}t}\ .
	\label{eq:pid}
\end{equation}

Here, $u(t)$ is the control variable and $e(t) = r(t) - y(t)$ is the error given by the difference between the SP ($r(t)$) and PV ($y(t)$). $x$ is a bound variable within the integral. In practice, in digital PID controllers, this continuous PID equation is discretised and the integral and derivative are approximated by finite difference methods. \cite{visioli_practical_2006}

The PID controller can be augmented by a feed-forward action, which does not incorporate feedback, but tries to (partially) control the plant by using some pre-known model of it. An example of this is simply including the setpoint multiplied by a feed-forward gain $K_{\rm ff}$:

\begin{equation}
	u(t) = K_{\rm ff}\ r(t) + K_{\rm p}\ e(t) + K_{\rm i} \int_0^t e(x)\ \mathrm{d}x + K_{\rm d} \frac{\mathrm{d}e(t)}{\mathrm{d}t}\ .
	\label{eq:pid_ff}
\end{equation}

\subsubsection{Controller design}
In the first version of the humidistat \cite{veldscholte_design_2021}, a single control loop exists: the humidity controller. It directly regulates the humidity in the chamber by actuating the two valves as FCEs (\autoref{fig:pid_1}). While this works reasonably well after careful tuning of the PID gains, it suffers from lack of robustness. Most notably, it is sensitive to the feed pressure, and finding an universal `sweet spot' for the PID gains is difficult~\cite{veldscholte_design_2021}.

The main reason for the challenges in tuning the controller stems from imperfect actuators. The proportional solenoid valve response is non-linear and also exhibits rather severe hysteresis (from plunger stiction). This adversely affects the controller performance in several ways: first of all, the non-linearity results in a operating point-dependent process gain. Hence, the optimal PID gains for moderate humidity values are different from those for extreme humidity values. For example, because of the S-shaped flow rate-CV curve, a tuning parameter set that provides good performance for extreme humidity values might cause unstable\slash underdamped behaviour around moderate values. Moreover, the total flow rate is only constant if the valve response is linear. Non-linearities in the valve response cause variations in the total flow rate with CV, which in turn alter the process dynamics since the plant's response time and dead time depend on the total flow rate. Finally, hysteresis in the valve response can induce oscillation of the controller~\cite{di_capaci_augmented_2018}.

An elegant way to improve the dynamic performance and stability of a PID controller is to employ cascade control (\autoref{fig:pid_2}). In that case, there is one outer control loop which controls humidity, and two inner loops that control flow: one for each valve. The outer controller does not directly actuate the valves, but its output is fed into the (flow rate) setpoints of the two inner control loops~\cite{kuphaldt_lessons_2017, visioli_practical_2006}. As such, the inner control loops essentially constitute custom flow controllers. Naturally, such a setup requires flow sensors besides just a humidity sensor.

A cascade controller can improve dynamic performance by exploiting a separation of characteristic timescales in the plant: the inner loops, controlling flow, have much faster dynamics than the outer loop because the inner PVs (flow rates) respond much quicker to changes than the outer PV (chamber humidity)~\cite{kuphaldt_lessons_2017, visioli_practical_2006}. Moreover, a cascade controller is better able to deal with hysteresis in the plant \cite{li_frequency_2014}.

Another advantage of using a cascade controller is that the actuator non-linearities are much less of a concern since they are dealt with in the inner loop~\cite{visioli_practical_2006}; the flow rates are closed-loop controlled. As a result, the total flow rate can be kept constant despite non-linear valve response~\cite{kuphaldt_lessons_2017}. This is difficult to achieve without closed-loop flow rate control: one could try to compensate for the non-linear valve response, but that requires carefully characterising the flow rate-CV curve, which is complicated by hysteresis and dependence on external factors such as the feed pressure. Closed-loop control also removes the need for carefully calibrating the zero-flow solenoid duty cycle.
% ^ duidelijk genoeg / beter verwoorden?

\subsection{Components}
\subsubsection{Microcontroller board}
As a microcontroller board, a Teensy LC~\cite{teensylc} is used, which is similar in terms of price to the more popular Arduino Uno which was used in the previous version, but features much improved specifications: it is faster, has more flash and RAM and more I/O pins. This allows for a more complex UI. A comparison between the specifications of the Teensy LC and Arduino Uno is given in \autoref{tab:uno_teensy}.

OpenHumidistat can be used fully standalone, but it is possible to connect it to a PC to monitor and\slash or log the controller parameters using a Python utility.

\begin{table}[h]
	\centering
	\caption{Comparison between specifications of the Arduino Uno and Teensy LC.}
	\begin{tabular}{@{}llll@{}}
		\toprule
		\multicolumn{2}{l}{}                 & Arduino Uno & Teensy LC          \\ \midrule
		\multicolumn{2}{@{}l}{Price}         & \begin{tabular}[c]{@{}l@{}}€20 (official)\\ €5 (compat.)\end{tabular} & €13 \\[0.3em]
		\multicolumn{2}{@{}l}{MCU} & ATmega328P  & \begin{tabular}[c]{@{}l@{}}ARM\\Cortex-M0+\end{tabular}\\[0.3em]
		\multicolumn{2}{@{}l}{Clock}     & 16 MHz      & 48 MHz          \\[0.3em]
		\multirow{2}{*}{Memory}& Flash       & 32 kB       & 62 kB              \\
		& SRAM        & 2 kB        & 8 kB               \\[0.3em]
		Voltage      &             & 5 V         & 3.3 V              \\[0.3em]
		\multicolumn{2}{@{}l}{Digital I/O pins} & 14          & 27              \\[0.3em]
		\multirow{2}{*}{ADC}   & Channels    & 6           & 13                 \\
		& Resolution  & 10 bit      & 12 bit             \\[0.3em]
		\multirow{2}{*}{PWM}   & Channels    & 6           & 10                 \\
		& Resolution  & 8 bit       & 16 bit\\
		\bottomrule
	\end{tabular}
	\label{tab:uno_teensy}
\end{table}

\subsubsection{UI}
The 16x2 character display from the previous version has been upgraded to a larger, graphical 128x64 LCD.

Meanwhile, similarly to the previous version, a 5-button keypad with 4 directional and a `select' button is used. This keypad is implemented as a resistance ladder, such that it only occupies a single analog input pin on the microcontroller.

If desired, this keypad could relatively simply be interchanged for an alternative input method, such as a joystick or rotary encoder.

\subsubsection{Sensors}
The humidity sensor was upgraded from the DHT\-22\slash AM2302 to the more reliable and accurate Sensirion SHT-85~\cite{smith_compare_2017}. This sensor, although more expensive, is also significantly smaller, which makes it easier to fit to measurement chambers.

In order to provide flow rate feedback for closed-loop flow control, MEMS flow sensors from Omron (model D6F-P0010A1) were used. These sensors output the flow rate as an analog voltage signal, and have a maximum measurable flow rate of \SI{1}{\liter\per\minute}.

\subsubsection{Valves}
In order to achieve analog control over the flow rates, proportional solenoid valves were used. Unchanged from the previous version, PVQ30-series miniature proportional solenoid valves from SMC were used.

These valves are driven by a (smoothed) PWM current that is generated by a custom solenoid valve driver (see \ref{sec:circuits}).

\subsubsection{Humidifier}
One of the two air streams is humidified by bubbling it through water in two gas washing bottles. Two bottles are connected in series to ensure sufficient water vapour saturation at high flow rates.

This method is simple and does not require any electronics. Moreover, unlike ultrasonic humidifiers, it strictly produces vapour, and no aerosols.

\subsection{Summary}
In summary, \name{} features:
\begin{itemize}
    \item Standalone control of humidity in a chamber
    \item Optional monitoring\slash logging of controller state using Python utility on a PC connected over USB
    \item High accuracy and good disturbance rejection through closed-loop control
    \item Broad attainable humidity range of approximately \SIrange{10}{90}{\percent} (depending on humidifier efficacy and feed gas conditions)
    \item Easy and intuitive operation
    \item Excellent versatility and portability: can be applied to a wide range of experimental setups
\end{itemize}

\section{Design files}
\subsection{Design files summary}
\autoref{tab:designfiles_summary} contains a summary of the design files and links to the repositories they are stored in.
\begin{table*}[ht]
    \centering
    \caption{Summary of source files.}
    \begin{tabular}{@{}llll@{}}
        \toprule
        Name                  & Type & License & Location \\
        \midrule
        Hardware  & \begin{tabular}[c]{@{}l@{}}KiCAD project\\ (schematic, board layout)\end{tabular}  & CERN-OHL-S v2 & \href{https://dx.doi.org/10.17605/OSF.IO/UF3BN}{10.17605/OSF.IO/UF3BN} \\[0.5em]
        Firmware/software & \begin{tabular}[c]{@{}l@{}}Firmware: C++ code\\ Software: Python code\end{tabular} & GNU GPL v3 & \href{https://dx.doi.org/10.17605/OSF.IO/U3YNG}{10.17605/OSF.IO/U3YNG} \\[0.5em]
        Enclosure & \begin{tabular}[c]{@{}l@{}}3D model\\ (SolidWorks source, STL)\end{tabular}  & CERN-OHL-S v2 & \href{https://dx.doi.org/10.17605/OSF.IO/T49EH}{10.17605/OSF.IO/T49EH} \\
        \bottomrule
    \end{tabular}
	\label{tab:designfiles_summary}
\end{table*}

\begin{table*}[ht]
    \centering
    \caption{Summarised bill of materials.}
    \begin{tabular}{@{}llrr@{}}
        \toprule
        Qty & Item              & Unit price (€) & Ext. price (€) \\
        \midrule
        1 & Solenoid driver board (incl. components) & 1.67 & 1.67 \\
        1 & Mainboard (incl. components) & 0.35 & 0.35 \\
        1 & Teensy LC microcontroller board & 13.00 & 13.00 \\
        1 & 128x64 ST7920 LCD & 9.00 & 9.00 \\
        1 & Keypad & 0.75 & 0.75 \\
        1 & 12V 1A DC power supply & 8.00 & 8.00 \\
        1 & Humidity sensor (Sensirion SHT-85) & 26.35 & 26.35 \\
        2 & MEMS flow sensor (Omron D6F-P0010A1) & 45.38 & 90.76 \\
        2 & Proportional solenoid valve (SMC PVQ31-6G-16-01F) & 82.16 & 164.32 \\
        2 & Gas washing bottle & 70.00 & 140.00 \\
          & Cabling and connectors & 9.04 & 9.04 \\
          & Pneumatic tubing and couplings & 24.35 & 24.35 \\
        \midrule
        \multicolumn{3}{r}{Grand total} & 489.10 \\
        \bottomrule
    \end{tabular}
    \label{tab:bom_summary}
\end{table*}

\subsection{Electronic circuits}
\label{sec:circuits}

\begin{figure*}[p!]
	\centering
	\includegraphics[width=\linewidth]{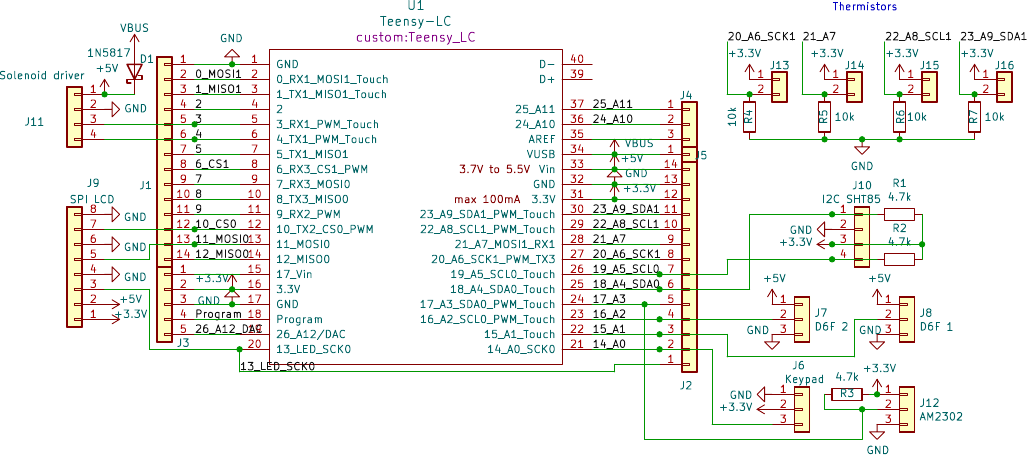}
	\caption{Circuit schematic for the mainboard.}
	\label{fig:schematic_mainboard}
\end{figure*}

\begin{figure*}[p!]
	\centering
	\includegraphics[width=\linewidth]{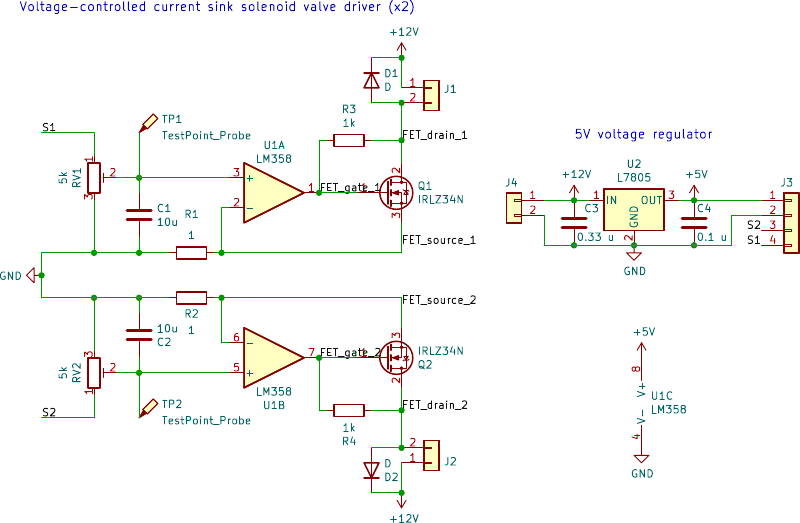}
	\caption{Circuit schematic for the solenoid driver.}
	\label{fig:schematic_solenoiddriver}
\end{figure*}

Electronic circuit schematics and PCB designs are provided for two boards: a mainboard and a solenoid driver board. Schematics of the circuits are shown in \autoref{fig:schematic_mainboard} and \ref{fig:schematic_solenoiddriver} respectively.

The microcontroller board (a Teensy LC) can be plugged onto the mainboard, which provides headers to connect to all peripherals (solenoid driver board, sensors, keypad, display). Besides headers, it contains a number of pull-up\slash down resistors and a Schottky diode between the USB \texttt{V\textsubscript{BUS}} and the device's \texttt{+5V} net, to avoid potentially backpowering the USB host in case the device is powered by its own power supply and USB simultaneously.

The solenoid driver board receives low-power PWM control signals from the microcontroller, and uses that to drive the (two) solenoid valves. More specifically, it comprises a 2-channel VCCS (Voltage-Controlled Current Sink), to drive the solenoid valves in a constant-current manner. This is necessary because the solenoids heat up considerably during normal operation, and this consequently alters their resistance. Without a constant-current driver, the solenoid current, which is the parameter that controls the flow rate, would drift with temperature. Design details about the VCCS are available in the supplementary material of the previous publication \cite{veldscholte_design_2021}.

The solenoids require \SI{12}{\volt}. This is provided by an external power supply, which is connected to header \texttt{J4} on the solenoid driver board. To allow the microcontroller and peripherals also to be powered through this power supply, the solenoid driver board includes a linear voltage regulator which creates a \texttt{+5V} line from the \texttt{+12V} supply.

An archive containing Gerber and drill files of the PCB designs, as well as source files of the circuit schematics and PCB designs which can be opened in the free EDA KiCad~\cite{noauthor_kicad_nodate} are made available under the CERN-OHL-S.

\begin{figure*}[ht]
    \centering
    \includegraphics[height=20em]{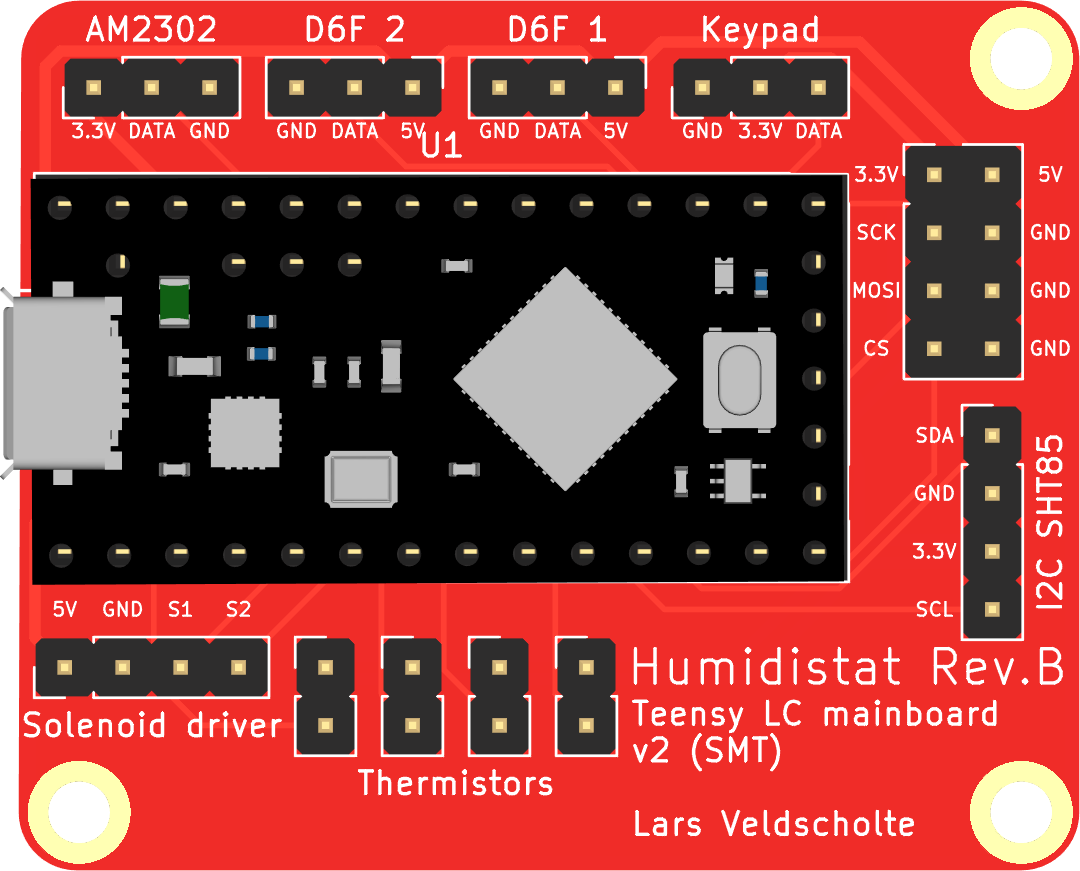}
    \hfill
    \includegraphics[height=20em]{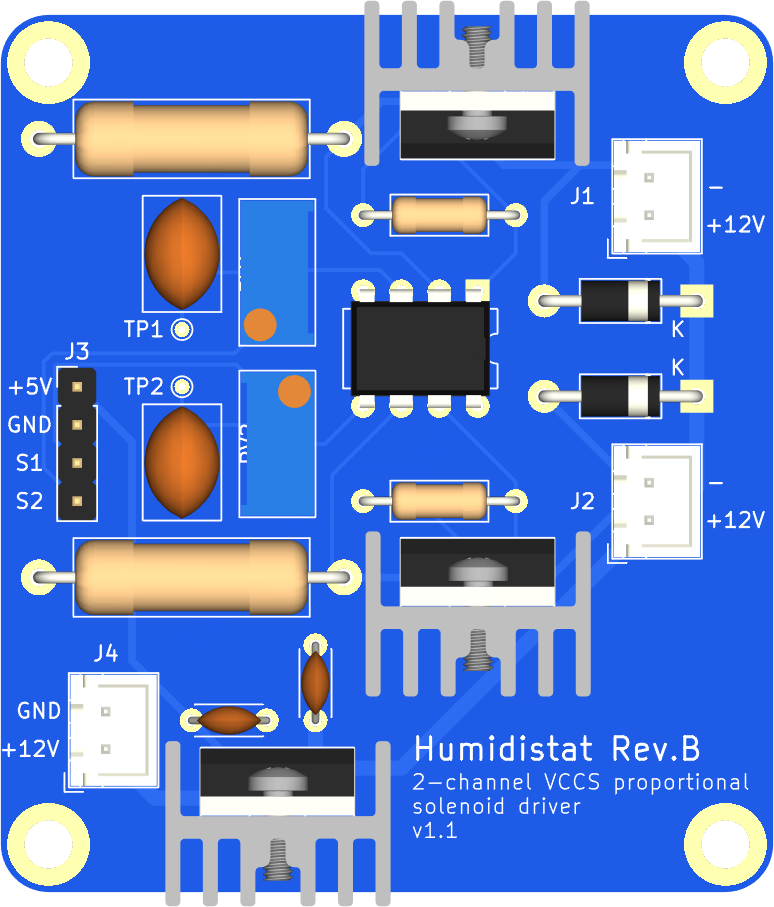}
    \caption{Renders of the two PCBs (not to scale).}
    \label{fig:pcb_renders}
\end{figure*}

\subsection{Firmware}
\label{sec:firmware}
The firmware running on the microcontroller is written in C++ with the Arduino Core framework and the PlatformIO~\cite{platformio} build system. It is written using the object-oriented programming paradigm, and as such consists of a number of modular classes.

The source is available under the GNU GPL v3.

\subsection{Enclosure}
CAD models for the enclosure are provided: source files (SolidWorks) and STL files for 3D printing.

\section{Bill of materials}
\label{sec:bom}
The full bill of materials is available at \url{https://dx.doi.org/10.17605/OSF.IO/F5U6E}.

We note that electronic components are listed with their prices per unit, but are commonly purchased in larger quantities (usually at least 10). This means that the actual cost might be inflated somewhat if one does not already have access to these (rather common) electronic components. Still, since these components are relatively cheap, even having to purchase 10 times the required amount does not increase the total price of the device by much.

Electronic and pneumatic components can be sourced from retailers such as RS Components, Mouser, Farnell, and Digikey. The gas washing bottles can be sourced from laboratory glassware distributors such as VWR and Fisher Scientific.

A summarised version of the BoM can be found in \autoref{tab:bom_summary}. It can be seen that the custom parts of the device are very cheap to build, and the majority of the total cost of the device comes from off-the-shelf components such as the humidity sensor, the flow sensors, gas washing bottles, and most notably the proportional solenoid valves. The total cost of the materials needed to build \name{} is less than €500.

\section{Build instructions}
\subsection{Materials and tools}
Refer to the BoM (section \ref{sec:bom}) for a comprehensive list of materials to order.

The PCBs can be ordered for quite cheap from various PCB manufacturers: at the cheapest fab houses, 5 units can be purchased for merely \$2 (excluding shipping). To do so, submit the Gerber and drill files (see section \ref{sec:circuits}) to a manufacturer of your choice. An online price comparison tool can be found at \url{https://pcbshopper.com}.

Building the device requires access to some basic tools. For the electronics, these are:

\begin{itemize}
    \item Soldering iron (+ solder)
    \item Wire cutters
    \item Stripping pliers
    \item Crimping tools for:
    \begin{itemize}
        \item DuPont connectors
        \item JST connectors
        \item Modular connectors
    \end{itemize}
    \item Tweezers
    \item Multimeter
\end{itemize}

For the enclosure, STL files are provided, intended to be 3D printed. Alternatively, the enclosure can be made in a different way, such as by modifying a universal enclosure. For this, one needs (access to) a workshop with some (machine) tools.

Additionally, some commonplace mounting materials (not listed on the BoM) are required, such as screws, nuts, and standoffs.

\subsection{Flashing the microcontroller}
\begin{enumerate}
    \item Configure the OpenHumidistat firmware by modifying \texttt{src\slash config.h} as required.
    \item Flash the OpenHumidistat firmware to the Teensy microcontroller by connecting it to a PC using a USB (Type-A to Micro-B) cable.
\end{enumerate}

Detailed instructions can be found in the \texttt{README\-.md} file included with the firmware (see section \ref{sec:firmware}).

\begin{figure*}[ht]
	\centering
	\begin{subfigure}[t]{0.5\textwidth}
		\centering
		\includegraphics[height=15em]{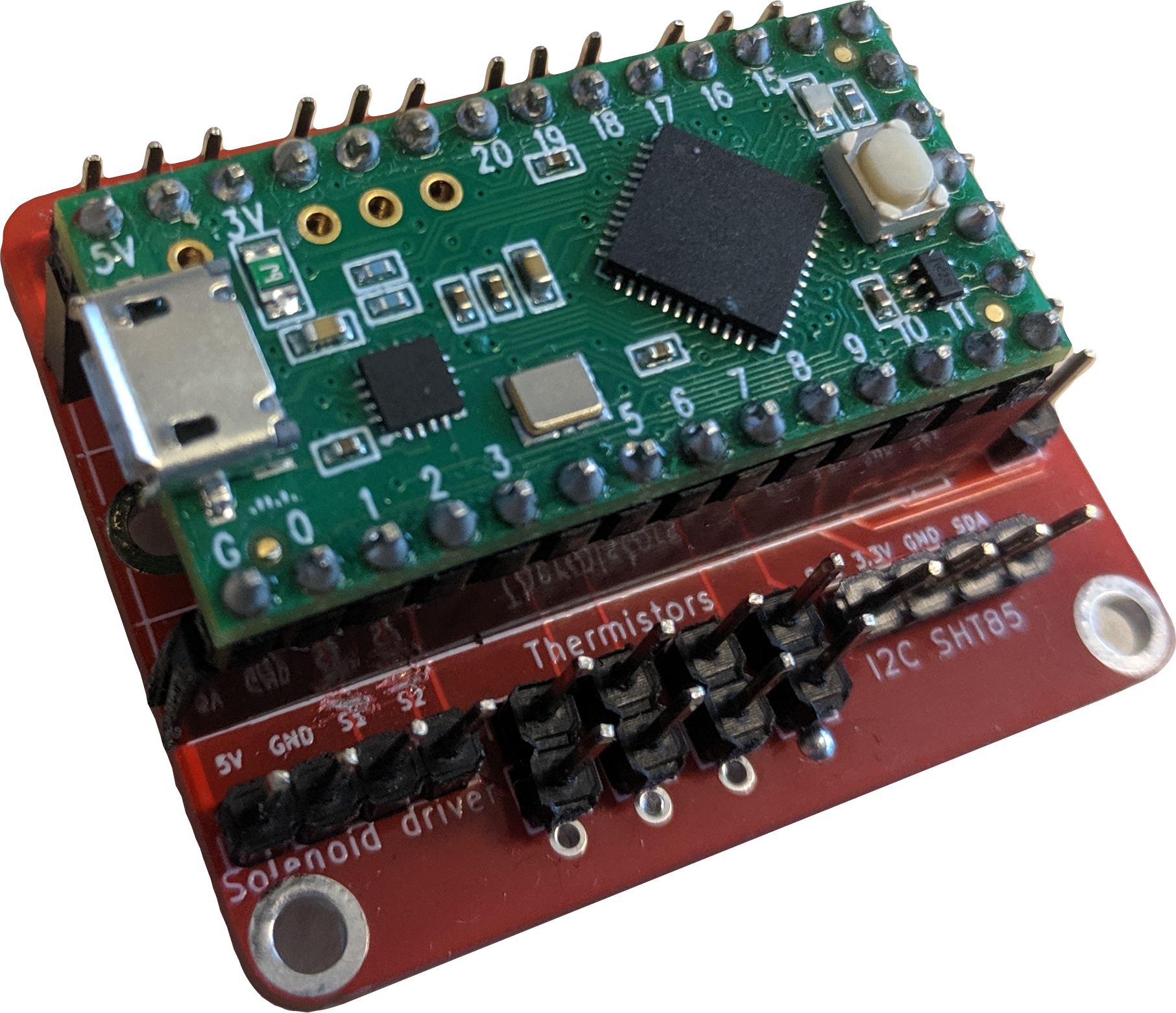}
		\caption{Assembled mainboard with Teensy installed.}
		\label{fig:picture_mainboard}
	\end{subfigure}%
	\begin{subfigure}[t]{0.5\textwidth}
		\centering
		\includegraphics[height=15em]{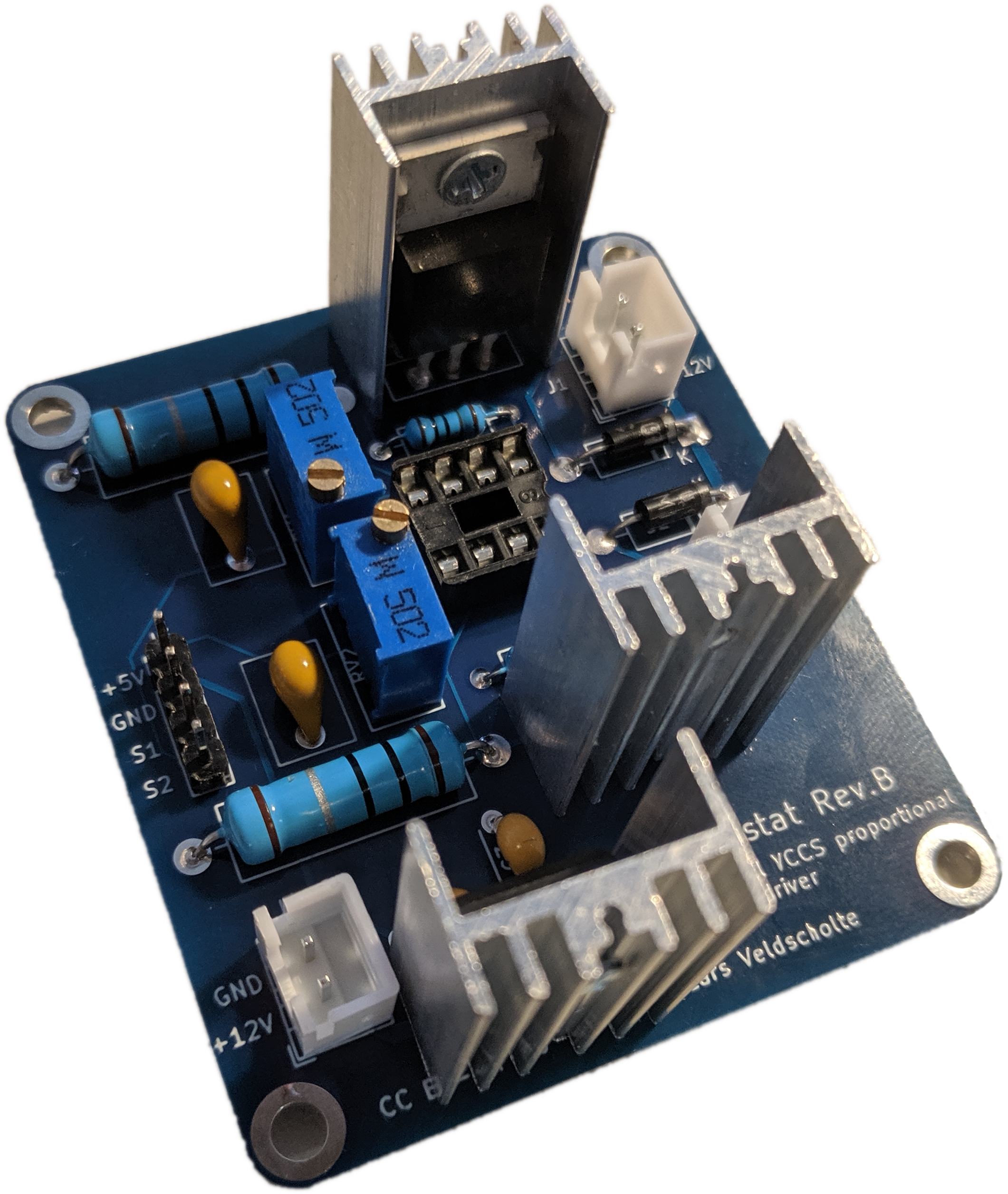}
		\caption{Assembled solenoid driver board (opamp IC not yet installed).}
		\label{fig:picture_solenoiddriver}
	\end{subfigure}
	\caption{}
\end{figure*}

\subsection{Electronics}
After obtaining the electronic components and the PCBs, the boards can be assembled. When populating the PCBs with the components, use the reference designators printed in silkscreen on the PCB and the BoM (or circuit schematic) to look up which component goes where. \autoref{fig:pcb_renders} can also be used as a visual reference.

\subsubsection{Solenoid driver board}
\begin{enumerate}
    \item The MOSFETs and the linear regulator used on the solenoid driver board (\texttt{Q1}, \texttt{Q2}, and \texttt{U2}) dissipate power as heat and should have heatsinks attached for cooling. Screw these on before soldering them onto the board.
    \item Populate the board with the components.
    \begin{itemize}
        \item It is recommended to start with the small components (resistors, diodes, disc capacitors), and solder the largest components (trimpots and TO-220 packages with heatsinks) last.
        \item For the DIP-8 IC (the LM358 opamp, \texttt{U1}), it is recommended to use a socket in order to prevent overheating the IC during soldering and to facilitate replacement if necessary.
    \end{itemize}
\end{enumerate}

When done, the board should look like in \autoref{fig:picture_solenoiddriver}.

\subsubsection{Mainboard}
The solenoid driver board is all THT (Through-Hole Technology) for easy hand-soldering, but the mainboard contains a few SMD (Surface Mount Device) components because of space constraints. However, it should still be doable to hand-solder these as long as it is done carefully and small-gauge solder wire is used.

\begin{enumerate}
    \item Start by disconnecting the Teensy's \texttt{+5V} and \texttt{V\textsubscript{BUS}} nets by cutting the trace between the two respective pads using a knife or sharp flat screwdriver, as described in the Teensy documentation~\cite{teensy_power} (Option \#1). The \texttt{V\textsubscript{BUS}} and \texttt{+5V} nets are connected through a Schottky diode on the mainboard.
    \item If the Teensy came without headers pre-installed, solder (male) headers to the microcontroller board.
    \item Solder the SMD components (7 resistors and a Schottky diode) onto the board. It will be a lot easier to do so before the THT components are in place.
    \begin{itemize}
        \item If the thermistors are not used, the resistors \texttt{R4}, \texttt{R5}, \texttt{R6}, and \texttt{R7} may be omitted.
    \end{itemize}
    \item Solder the rest of the components (headers and sockets) onto the board.
\end{enumerate}

When done, the board should look like in \autoref{fig:picture_mainboard}.

Cut off the component leads after they are secured in place with solder joints. Make sure all joints are sound and that there are no unintended solder bridges that could inadvertently short pads. Also make sure that all polarised components (diodes, transistors, ICs, connectors) are installed the right way around.

\subsection{Cables and connectors}
The several boards and peripherals are interconnected through (DuPont) jumper cables. Since these connectors are not keyed, it is good practise to plug the female DuPont connectors to their headers with their contacts facing out in order to not confuse the polarity of the connectors.

\subsubsection{Humidity sensor(s)}
\begin{enumerate}
    \item Create the humidity sensor cable: solder one end of a length (\SIrange{1}{2}{\meter} recommended, depending on the desired distance between the OpenHumidistat and the measurement chamber) of 4-wire data cable to the SHT-85 sensor and crimp a 6P4C plug on the other end. Mind the pin order: it should match that of the header marked \texttt{I2C SHT85} on the mainboard.
    \item Make a cable to connect the female modular jack (which will be mounted in the front panel) to the mainboard: crimp a 4-pin female DuPont connector on one end, and solder the wires on the other end to the pins of the female modular jack. Again, mind the order of the pins: the cable should preserve it along the entire path from the header on the mainboard, to the actual SHT-85 sensor. Use a multimeter to verify the continuity of the connections, if required.
\end{enumerate}

For backwards compatibility, the mainboard also includes a header for connecting a DHT22\slash AM2302 humidity sensor. If desired, the above procedure can be repeated for this sensor type.

\begin{figure*}[ht]
	\begin{subfigure}[t]{0.32\textwidth}
		\centering
		\includegraphics[height=13em]{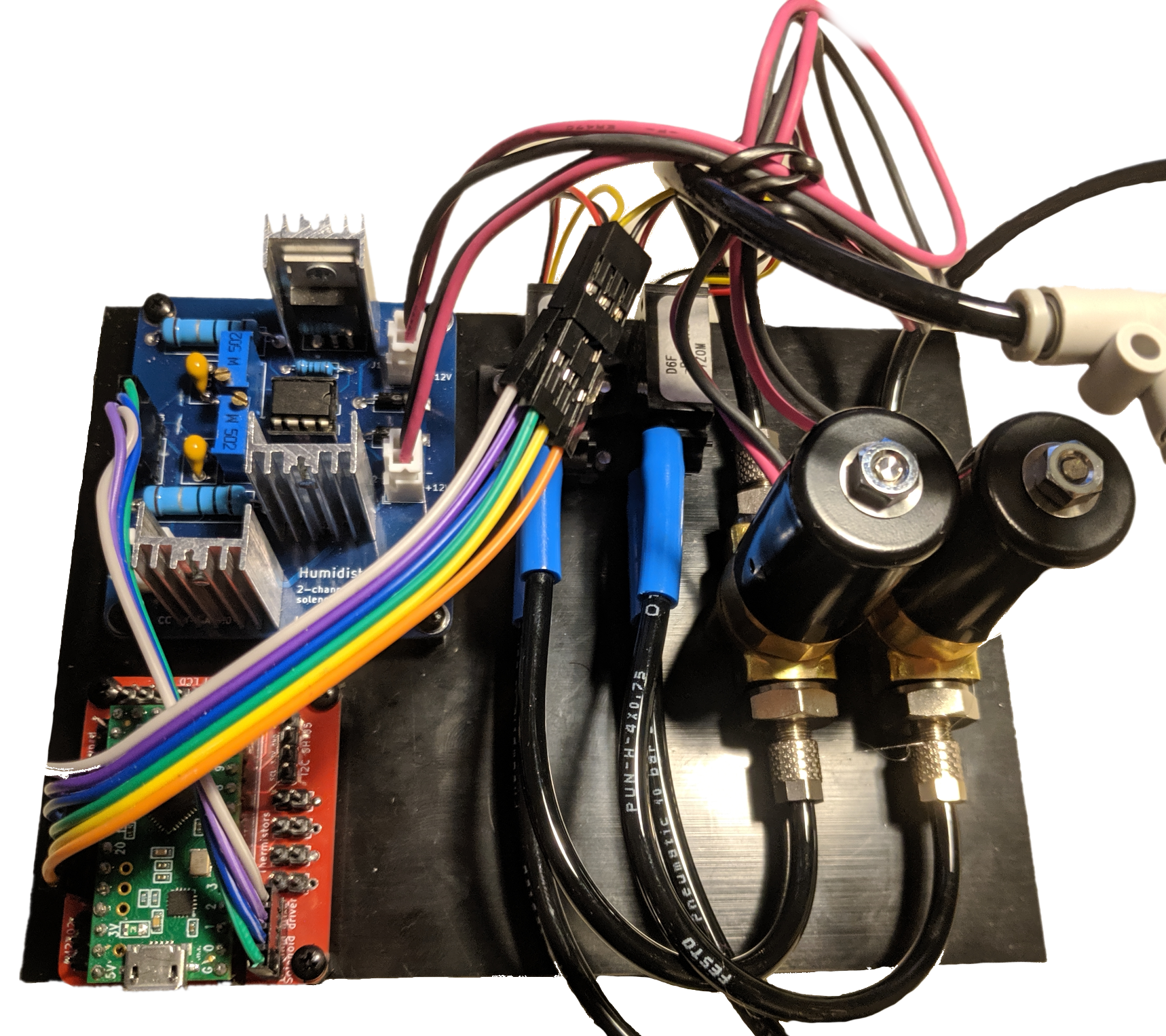}
		\caption{Baseplate with components installed and partially connected.}
	\end{subfigure}
	\hfill
	\begin{subfigure}[t]{0.32\textwidth}
		\centering
		\includegraphics[height=13em]{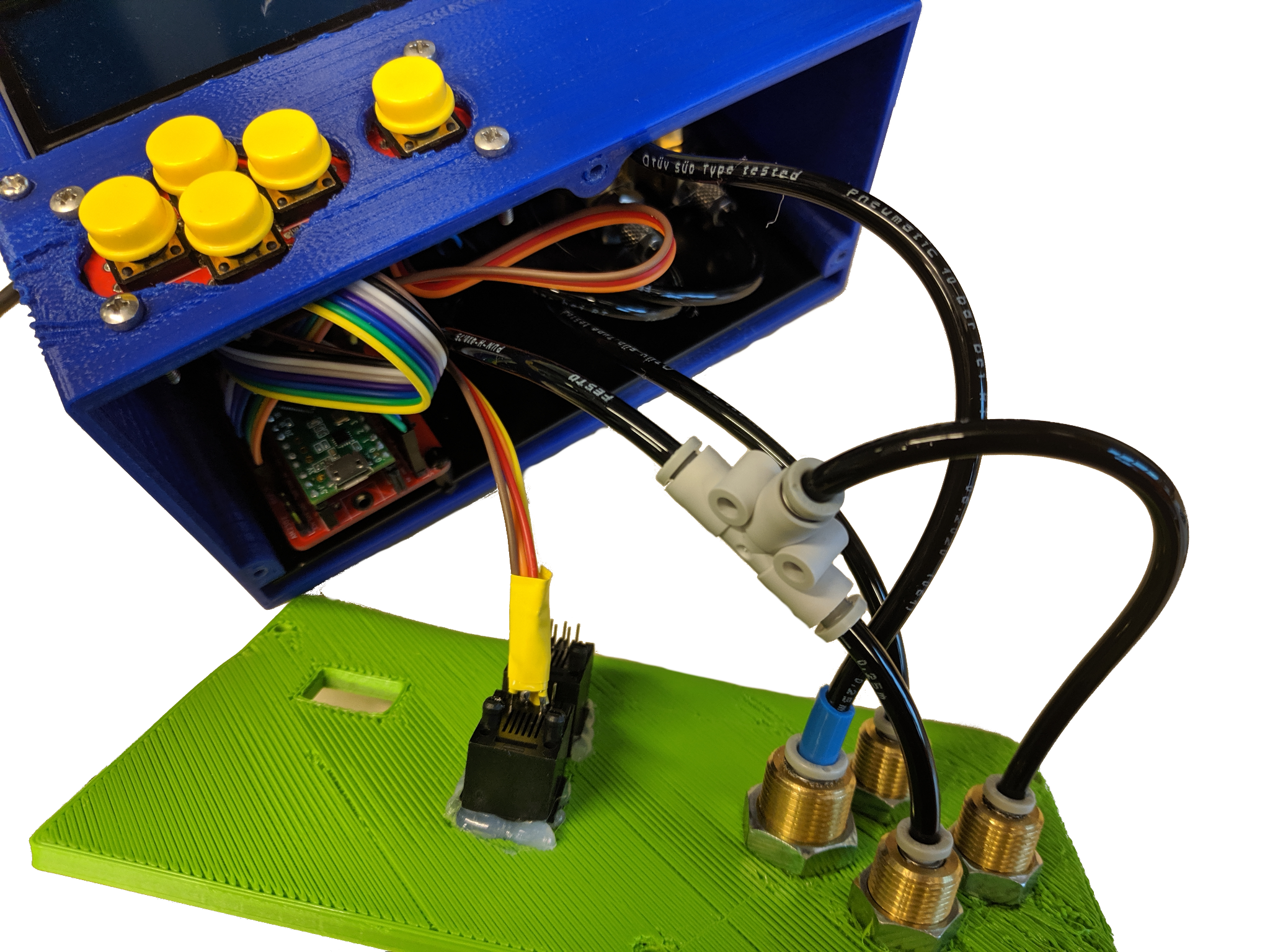}
		\caption{Baseplate installed in the 3D printed enclosure, with frontpanel open.}
	\end{subfigure}
	\hfill
	\begin{subfigure}[t]{0.32\textwidth}
		\centering
		\includegraphics[height=13em]{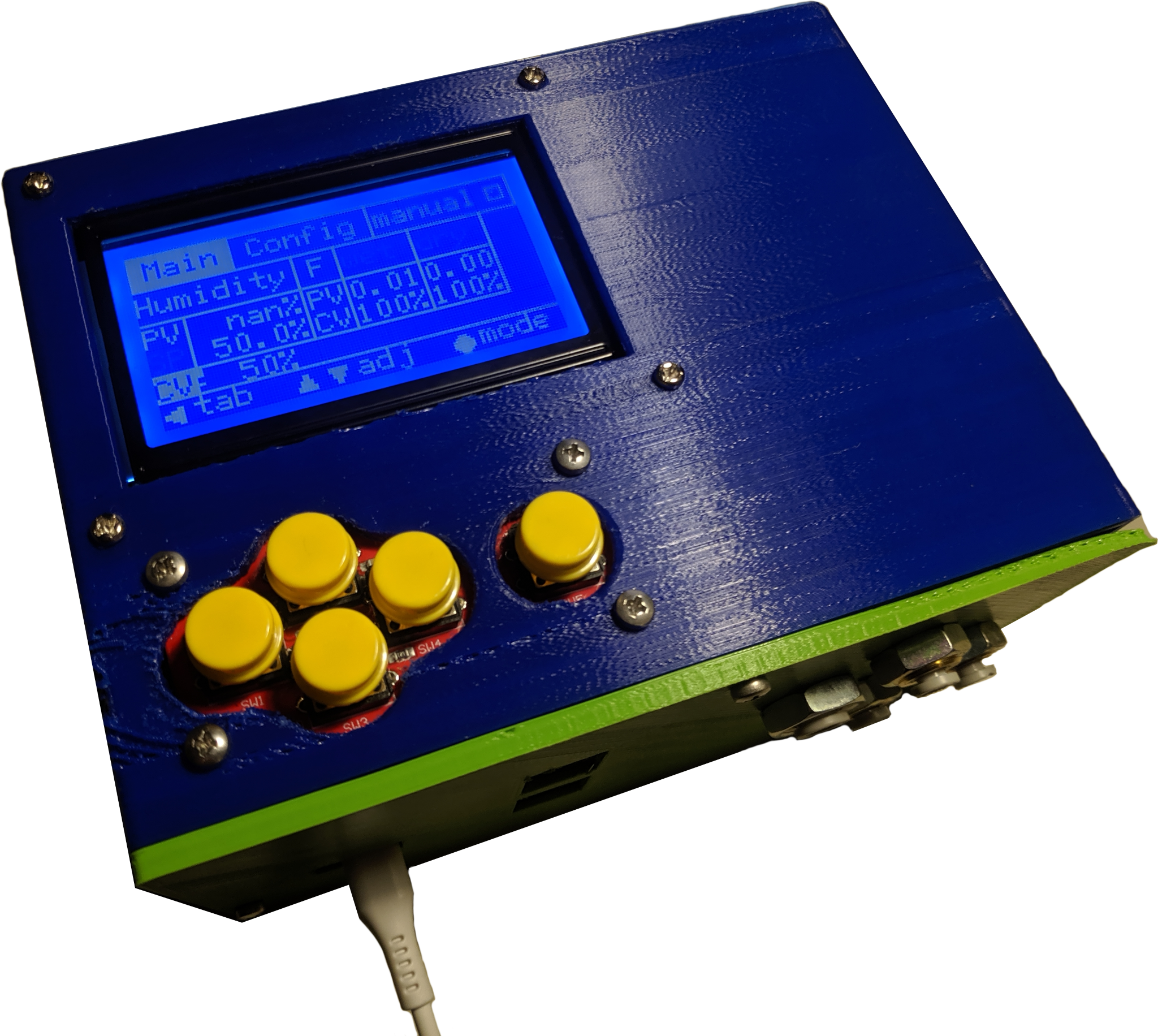}
		\caption{Finished device in 3D printed enclosure.}
	\end{subfigure}
	\caption{}
\end{figure*}

\subsubsection{Power}
Make a cable to connect the the barrel jack (which will be mounted in the front panel) to the \texttt{+12V} header (\texttt{J4}) on the solenoid driver board. For this, wire thicker than that used for the signal cables should be used: at least \SI{0.5}{\milli\meter\squared} (AWG 20).
\begin{enumerate}
    \item Crimp a JST-XH female connector to one end of two wires, and solder the other ends to the pins of the barrel jack. Mind the polarity: it should match that of the barrel plug of the power supply used. Usually, it is `center positive', but verify this: reverse polarity will damage the electronics.
    \item Use heat-shrink tubing to insulate the end of the barrel jack.
\end{enumerate}

\subsubsection{Solenoid valves}
\begin{enumerate}
    \item Crimp female 2-pin JST-XH connectors to the solenoid valve wires. Mind the polarity: the red wire should correspond to the positive bottom pin on headers \texttt{J1}\slash\texttt{J2} (marked \texttt{+12V}).
\end{enumerate}

\subsubsection{Thermistors}
Optionally, 4 thermistors can be installed for temperature monitoring of the solenoids and MOSFETs. For this,
\begin{enumerate}
    \item Crimp female 2-pin DuPont connectors to the thermistor cables.
    \item Glue or otherwise attach the thermistors to the solenoid valves and MOSFET heatsinks.
\end{enumerate}

\subsubsection{Other peripherals}
The other peripherals (display, keypad, flow sensors) are connected using (female-female) DuPont cables, which can be simply used as-is.

Refer to \autoref{tab:pins_display} for connecting the display module to the mainboard.

\begin{table}[ht]
    \centering
    \caption{Pin mapping between mainboard and ST7920 display module.}
    \begin{tabular}{@{}lll@{}}
        \toprule
        \multicolumn{2}{l}{Mainboard} & ST7920 display \\ \midrule
        \#           & Label          & Label         \\ \midrule
        1            & \texttt{+3.3V} & \texttt{BLA}  \\
        2            & \texttt{+5V}   & \texttt{VCC}  \\
        3            & \texttt{SCK}   & \texttt{E}    \\
        4            & \texttt{GND}   & \texttt{BLK}  \\
        5            & \texttt{MOSI}  & \texttt{RW}   \\
        6            & \texttt{GND}   & \texttt{PSB}  \\
        7            & \texttt{CS}    & \texttt{RS}   \\
        8            & \texttt{GND}   & \texttt{GND}  \\
        \bottomrule
    \end{tabular}
    \label{tab:pins_display}
\end{table}

\subsection{Testing and calibration}
\begin{enumerate}
    \item Partially assemble the device: install the Teensy onto the mainboard and connect the solenoid driver board, display, keypad, and flow sensors to the mainboard.
    \item When connecting the device over USB (connected to the Teensy), the device should power up. Check that the firmware is running, that the interface is shown on the display, and that you can interact with it using the keypad.
    \item Calibrate the solenoid driver by adjusting the trimpots. To do this:
    \begin{enumerate}
        \item Make sure there is a \SI{3.3}{\volt} signal applied to the pins \texttt{S1}\slash\texttt{S2} of the solenoid driver board. This is most straightforwardly done by having it connected to the mainboard, and having the microcontroller output a 100\% duty cycle signal on both of the channels. This should happen automatically, as the flow controllers' CVs will saturate with no actual flow through the sensors.
        \item After verifying on the display that the microcontroller is outputting a 100\% duty cycle signal on both channels, use a multimeter to measure the voltage between each of the test points \texttt{TP1} and \texttt{TP2}, and ground. Adjust each trimpot (\texttt{RV1} and \texttt{RV2}) until the voltage at the respective test point reads exactly \SI{0.330}{\volt}.
    \end{enumerate}
\end{enumerate}

\begin{figure*}[ht]
	\centering
	\includegraphics[width=0.9\linewidth]{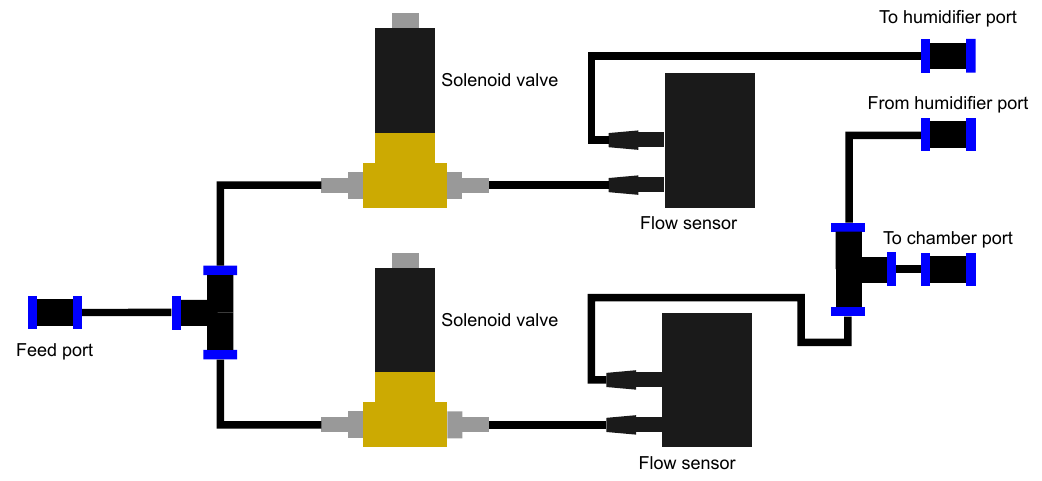}
	\caption{Diagram illustrating the pneumatic connections between the components.}
	\label{fig:pneumatics}
\end{figure*}

\subsection{Enclosure}
\begin{enumerate}
    \item Print the enclosure using the included STL files (two parts: the enclosure, and the frontpanel)
    \item Make a base plate onto which to affix the components.
    \begin{enumerate}
        \item Cut a piece of PVC to size.
        \item Drill holes in the base plate for affixing the components to it, and it to the bottom of the enclosure.
    \end{enumerate}
    \item Screw the components to the baseplate:
    \begin{enumerate}
    	\item The PCBs (use \SI{6}{\milli\meter} spacers)
    	\item The flow sensors
    	\item The solenoid valves
    \end{enumerate}
    \item Tap M3 thread in the holes in the enclosure for affixing the frontpanel.
    \item Install components into the enclosure:
    \begin{enumerate}
        \item Affix the display board using \SI{6}{\milli\meter} spacers.
        \item Affix the keypad. You may need to reposition the pin header.
        \item Affix the DC barrel jack.
    \end{enumerate}
    \item Install components into the frontpanel:
    \begin{enumerate}
        \item Glue the modular jacks into place.
        \item Affix the pneumatic bulkhead union couplings.
    \end{enumerate}
    \item You can then slide the baseplate including components into the enclosure, and screw the frontpanel on (after connecting cables and tubing, of course). See \ref{sec:assembling} for details on the pneumatic connections.
\end{enumerate}

\subsection{Measurement chamber}
A wide variety of measurement chambers can be used, depending on the envisioned application of the humidistat. The only requirements are that it must fit the humidity sensor, and that it should be reasonably closed; if there is too much exchange with ambient air, it might be difficult to reach extreme humidity values. The device is tested with chambers with volumes of several \si{\centi\liter}, but smaller or larger chambers can be used if desired. Note that larger chambers will have larger residence times, which will in turn limit the speed of the controller.

If for a particular experimental setup it is undesirable or impossible to fit the humidity sensor in the measurement chamber, it also possible to construct a `pre-chamber' containing just the humidity sensor, which is then connected between the humidistat and the measurement chamber. This pre-chamber should be completely air-tight in this case. The advantage of this configuration is the versatility: the humidistat can be used with practically every experimental setup featuring a flow cell this way without any modification. A potential disadvantage arises when the actual measurement chamber is too open, in which case exchange with the ambient air occurs and the actual humidity will deviate from the measured humidity in the pre-chamber.

\subsection{Assembling the device}
\label{sec:assembling}
Once all the individual assemblies are completed, the full device can be assembled.

\begin{enumerate}
    \item Screw the G1/8 <--> \SI{4}{\milli\meter} barbed fittings into the solenoid valve ports.
    \item Make the pneumatic connections using PU tubing. See \autoref{fig:pneumatics} for reference.
\end{enumerate}

\section{Operation instructions}
To turn on the device, connect the external \SI{12}{\volt} power supply to the DC jack. The microcontroller (and peripherals) can also be powered from the USB port, but for operation of the valves, a \SI{12}{\volt} supply is required.

Instruction for the usage of the UI can be found in the \texttt{README.md} included with the firmware.

\subsection{Setting up the device}
\subsubsection{Humidifier}
\begin{enumerate}
    \item Fill the gas washing bottles with (DI) water.
    \item Connect the two gas washing bottles in series.
    \item Connect the gas washing bottles to the designated ports on the device. Make sure the direction is correct: if it is wrong, it will flush water into the system.
\end{enumerate}

\subsubsection{Chamber}
\begin{enumerate}
    \item Using PU tubing, connect the outlet port of the device to the measurement chamber.
    \item Also connect the humidity sensor installed in this measurement chamber to the device.
\end{enumerate}

\subsubsection{Feed}
As supply gas, either dry air or nitrogen can be used. Although not strictly required, the solenoid valves perform better when it is pressure-regulated (at \SIrange{1}{2}{\bar}).

\begin{enumerate}
    \item Connect the nitrogen\slash air supply to the inlet port of the device.
    \item Increase the feed pressure until the OpenHumidistat's flow controllers achieve sufficient flow (recognisable by the stopping of blinking of the flow controller labels on the display). Note that you need at least about \SI{1}{\bar} for the solenoid valves to open.
\end{enumerate}

\subsubsection{Testing and tuning}
The system contains two solenoid valves and two flow sensors: one set for humid flow, and one for dry flow. Hence, there are 4 ways to connect these, but they must be connected in the right configuration for the controllers to work properly. To test this:

\begin{enumerate}
    \item Turn on the device.
    \item With the humidity controller in manual mode and feed pressure applied to the device's inlet:
    \begin{itemize}
        \item If the flow controllers manage to successfully regulate the flow rates, the flow sensors correctly correspond to the solenoid valves.
        \item If both flow controllers saturate (one CV goes to 0\%, while the other goes to 100\%), the flow controllers are erroneously trying to control each other's flow. In this case, swap either:
        \begin{itemize}
            \item the pneumatic connections between the flow sensors and the solenoid valves,
            \item the connectors of either the flow sensors (on the mainboard) or the solenoid valves (on the solenoid driver board),
            \item or the pin numbers of either the flow sensors or the solenoid driver channels in the firmware's config file.
        \end{itemize}
    \end{itemize}
    \item Turn the humidity controller into automatic mode.
    \begin{itemize}
        \item If the humidity controller manages to regulate the humidity by converging on the SP (even if doing so slowly, or with overshoot), the device is configured correctly.
        \item If the humidity controller saturates (the CV goes to either 0\% or 100\% and stays there), the humid and dry flows are swapped. In this case, swap either:
        \begin{itemize}
            \item the pin numbers of \emph{both} the flow sensors and solenoid driver channels in the firmware's config file,
            \item the electrical connectors of \emph{both} the flow sensors and the solenoid valves
            \item or the pneumatic connections of the dry and wet flows (disconnect the tubing from the flow sensor going to the humidifier and connect it directly to the union tee, and disconnect the tubing going directly going to the union tee and connect it to bulkhead union going to the humidifier).
        \end{itemize}
    \end{itemize}
    \item Tune the two PID controllers (outer humidity controller and inner flow controllers) appropriately for the total flow rate and measurement chamber used by adjusting the controller parameters. Defaults for these can be set in \texttt{config.h}, but they can also be overridden using the UI itself in a standalone manner. Section \ref{sec:tuning} illustrates the tuning procedure.
\end{enumerate}

\subsection{Hazards}
Because the outlet of the gas washing bottles is subject to back pressure from the dry flow, there exists a risk of backflow through the gas washing bottles when the inlet of the gas washing bottles is left unconnected, or when the humid solenoid valve is closed (open) and the tubing and\slash or connections from there to the gas washing bottles are not completely air-tight. This is problematic because it will siphon liquid water out the gas washing bottles into the system.

Users should be alert of this potential issue, which can be spotted by the water level rising in the tubes of the gas washing bottles. If this problem occurs, it is strongly recommended to install a check valve (e.g. SMC AKH, Festo H-QS) in line with the gas washing bottles, which should prevent any backflow.

\section{Validation and characterisation}
\subsection{Tuning}
\label{sec:tuning}

\begin{figure*}[ht]
	\centering
	\begin{subfigure}[t]{0.5\textwidth}
		\includegraphics[width=\linewidth]{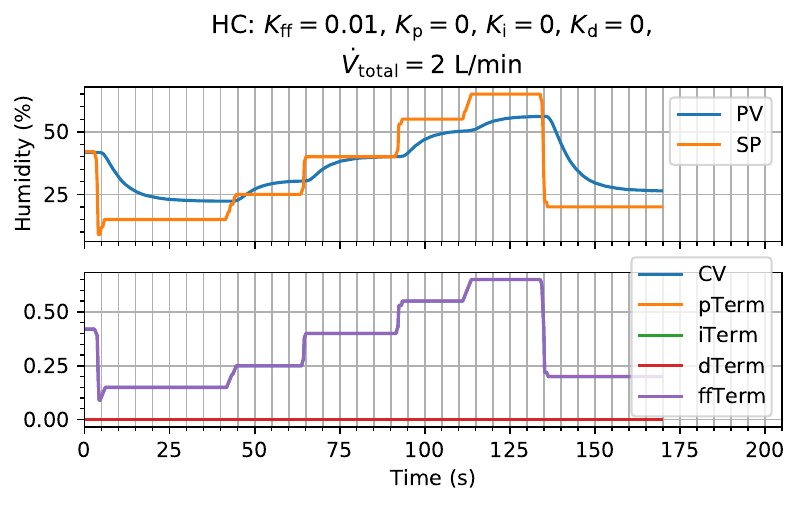}
		\caption{Controller response with feed-forward only.}
		\label{fig:tuning_example_ff}
	\end{subfigure}%
	\begin{subfigure}[t]{0.5\textwidth}
		\includegraphics[width=\linewidth]{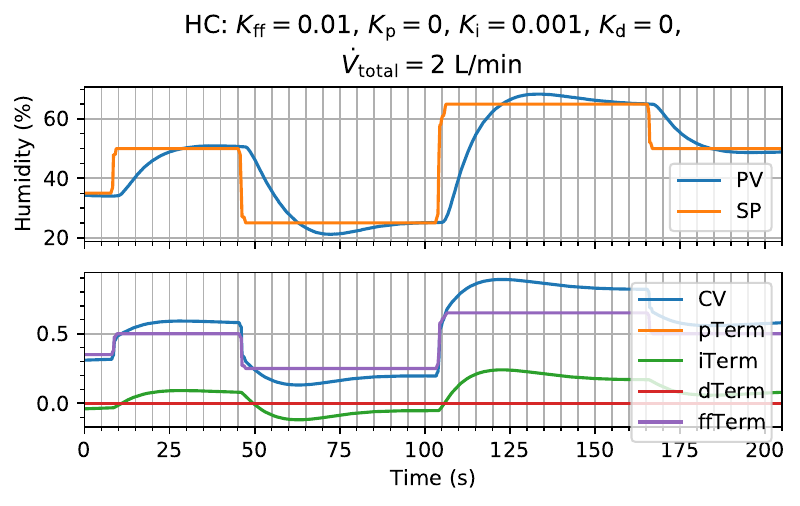}
		\caption{Controller response with feed-forward and integral action.}
		\label{fig:tuning_example_ff_i}
	\end{subfigure}
	\caption{}
\end{figure*}

A manual tuning procedure is outlined below. This was performed with a `universal pre-chamber' with a volume of \SI{25}{\milli\liter}, which with the used total flow rate of \SI{2}{\liter\per\minute} gives a residence time of \SI{0.75}{\second}.

\subsubsection{Flow controllers}
The inner control loops, constituting flow controllers, are tuned as PI-only controllers with a high $K_{\rm i}$, so that they essentially are integral controllers. This works very well because the plant, that is, the flow though the solenoid valve, is a fully self-regulating process that, although non-linear and exhibiting hysteresis, fortunately almost does not have any dead or response time: it responds virtually instantly to changes that the controller makes by actuating the valves. Hence, overshoot is barely of concern and this controller is very stable. Because the signal from the flow sensors is rather noisy and because it is not necessary in this case, derivative action is not used ($K_{\rm d} = 0$).~\cite{kuphaldt_lessons_2017}

As mentioned in the first section, a major improvement over the single-loop controller is that it is no longer critical to meticulously calibrate the minimum value of the CV (solenoid duty cycle). Whereas this was crucial in the previous design to make sure the resulting total combined flow rate is reasonably constant and no deadband exists at the lower range of the solenoid response, with closed-loop flow control the this zero-flow duty cycle is automatically `found' by the flow controller. In the cascade control scheme this parameter still exists, but solely for the proper functioning of the anti integral windup mechanism.

\begin{table}[hbt]
    \centering
    \caption{Tuning parameters}
    \begin{tabular}{@{}lllll@{}}
        \toprule
        & $K_{\rm ff}$ & $K_{\rm p}$ & $K_{\rm i}$ (\si{\per\second}) & $K_{\rm d}$ (\si{\second}) \\ \midrule
        FC & 0 & 0.005 & 0.05 & 0 \\
        HC & 0.01 & 0.025 & 0.002 & 0.025 \\
        \bottomrule
    \end{tabular}
    \label{tab:tunings}
\end{table}

\subsubsection{Humidity controller}
For the (outer) humidity controller, is helpful to start off by considering a purely \emph{feed-forward} (ff) controller. In that case, the controller solely acts on the SP, and there is no \emph{feedback}. In other words, the CV is purely a function of the SP.

Feed-forward-only controllers like these are rarely ever sufficient because they cannot deal with external disturbances (after all, if they were, the entire PID controller would be unnecessary), but including a significant feed-forward action in our outer humidity controller turns out to be a useful starting point. We set $K_{\rm ff}$ to $0.01$, since that creates a CV of $1$ (the maximum) for a SP of $100$ (\%)\footnote{In the PID code, the CV is normalised between 0 and 1, but the humidity PV/SP runs from 0 to 100.}. The implicit model behind this feed-forward action is that the resulting humidity of the mixed-together dry and humid air is proportional to the ratio of those two flow rates. This is a reasonable initial guess that nevertheless fails when the `dry' air is slightly moist, the humid air is not fully saturated, or when the plant is confronted with other external disturbances, but that does not matter, since the feedback action (PID) exists to correct for those. The main benefit of partly relying on feed-forward action, is that this can supplant most of the integral action in providing the `offset', yet without the potential for destabilising the controller.

\begin{figure*}[ht]
	\centering
	\begin{subfigure}[t]{0.5\textwidth}
		\centering
		\includegraphics[width=\linewidth]{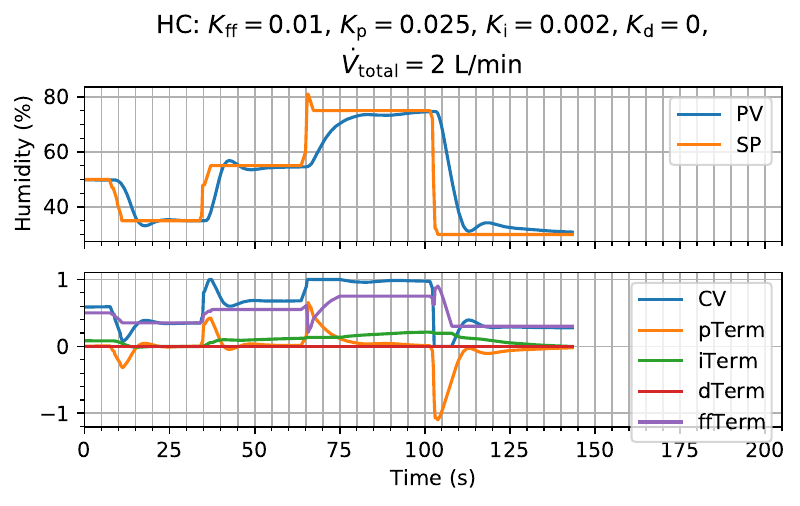}
		\caption{Controller response with feed-forward,\\ proportional, and integral action.}
		\label{fig:tuning_example_ff_i_p}
	\end{subfigure}%
	\begin{subfigure}[t]{0.5\textwidth}
		\centering
		\includegraphics[width=\linewidth]{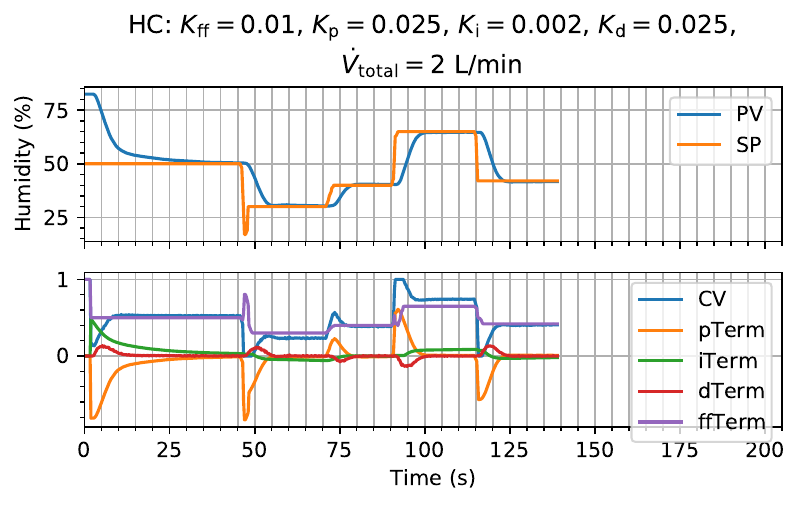}
		\caption{Controller response with feed-forward,\\ proportional, integral, and derivative action.}
		\label{fig:tuning_example_ff_i_p_d}
	\end{subfigure}
	\caption{}
\end{figure*}

The dynamic response of the feed-forward-only controller is shown in \autoref{fig:tuning_example_ff}. Although it is physically impossible for this controller to oscillate or overshoot, the PV values it settles on deviate significantly from the SP (for most values) and the settling times are far from optimal.

As an example, a small amount of integral action is then added to help with static deviations from the aforementioned ideal model. The resulting response is shown in \autoref{fig:tuning_example_ff_i}. The controller now eventually homes in on the SP, but only rather slowly, and after considerably overshooting it. This is an inherent effect of high integral action on a plant with some time delay~\cite{kuphaldt_lessons_2017, wescott2018}.

Next, we add proportional action. This makes the controller act briefly on the error between the PV and SP, and has the effect of reducing the accumulation time of the integral action, so the settling times improve (\autoref{fig:tuning_example_ff_i_p}).

Derivative action is also added to accelerate the dynamic response by reducing overshoot, and finally the response is reasonably satisfactory, with good setpoint tracking, and settling times on the order of \SI{10}{\second} (\autoref{fig:tuning_example_ff_i_p_d}).

\subsection{Robustness}
Due to closed-loop flow control, the system is a lot more robust to various feed pressures. Whereas with the single-loop humidistat variations in feed pressure and\slash or chamber pressure drop led to varying flow rates, which in turn affects the solenoid valve response and chamber response time, the flow controllers in the cascade humidity controller always keep the total flow rate constant despite varying feed pressures. This makes the controller more portable, especially in situations where the feed pressure is not controlled.

\subsection{Capabilities}
\begin{table}[H]
    \centering
    \begin{tabular}{@{}ll@{}}
    \toprule
    Range                   & 10 \% - 90 \%                    \\
    Maximum total flow rate & \SI{2}{\liter\per\minute}        \\
    Settling time           & \textasciitilde \SI{10}{\second} \\
    Accuracy                & ±1.5 \%                          \\
    Precision               & 0.01 \%                          \\
    \bottomrule
    \end{tabular}
\end{table}

\section{Conclusion and outlook}
We presented the design of a free and open-source humidistat. Due to the usage of cascade PID control, in which the dry and humid flow rates are also closed-loop controlled, the dynamic performance and robustness was significantly improved over our previous design in which a single control loop was used: Settling times are improved from \textasciitilde\ \SI{30}{\second} to \textasciitilde\ \SI{10}{\second}.

The versatility of the device makes it broadly applicable, bringing humidity control to a wide variety of experimental and analytical setups.

As a final note, we want to mention the possibility of generalising the OpenHumidistat beyond water vapour. By swapping the water in the gas washing bottles for a different substance, this design could readily function as a vapour pressure controller for arbitrary vapours, provided that an appropriate vapour sensor for it exists.

\section*{Author contributions}
\begin{description}
    \item[Lars B. Veldscholte] Conceptualisation, software, validation, investigation, writing - original draft, writing - review \& editing, visualisation
    \item[Sissi de Beer] Conceptualisation, supervision, funding acquisition, writing - review \& editing
\end{description}

\section*{Declaration of competing interest}
The authors declare no competing interests.

\section*{Acknowledgements}
The authors extend their thanks to Koen Jorissen for his help with CAD modelling, 3D printing, and assembling of the enclosure.

This research received funding from the Dutch Research Council (NWO) in the framework of the ENW PPP Fund for the top sectors and from the Ministry of Economic Affairs in the framework of the ‘PPS-Toeslagregeling’ regarding the Soft Advanced Materials consortium.

\onecolumn
\section*{List of abbreviations and symbols}
\begin{center}
	\begin{tabulary}{0.7\textwidth}{LL}
		\toprule
		PID & \textbf{P}roportional, \textbf{i}ntegral, \textbf{d}erivative (control) \\
		PWM & \textbf{P}ulse-\textbf{w}idth \textbf{m}odulation \\
		PCB & \textbf{P}rinted \textbf{c}ircuit \textbf{b}oard \\
		PV & \textbf{P}rocess \textbf{v}ariable \\
		SP & \textbf{S}et\textbf{p}oint \\
		CV & \textbf{C}ontrol \textbf{v}ariable \\
		FCE & \textbf{F}inal \textbf{c}ontrol \textbf{e}lement \\
		FC & \textbf{F}low \textbf{c}ontroller \\
		HC & \textbf{H}umidity \textbf{c}ontroller \\
		$K_{\rm p}$ & Proportional gain \\
		$K_{\rm i}$ & Integral gain \\
		$K_{\rm d}$ & Derivative gain \\
		$K_{\rm ff}$ & Feed-forward gain \\
		$\dot{V}$ & (Volumetric) flowrate \\
		UI & \textbf{U}ser \textbf{i}nterface \\
		I/O & \textbf{I}nput/\textbf{o}utput \\
		RAM & \textbf{R}andom-\textbf{a}ccess \textbf{m}emory \\
		ADC & \textbf{A}nalog-to-\textbf{d}igital \textbf{c}onverter \\
		VCCS & \textbf{V}oltage-\textbf{c}ontrolled \textbf{c}urrent \textbf{s}ink \\
		EDA & \textbf{E}lectronic \textbf{d}esign \textbf{a}utomation \\
		MOSFET & \textbf{M}etal–\textbf{o}xide–\textbf{s}emiconductor \textbf{f}ield-\textbf{e}ffect \textbf{t}ransistor \\
		IC & \textbf{I}ntegrated \textbf{c}ircuit \\
		THT & \textbf{T}hrough-\textbf{h}ole \textbf{t}echnology \\
		SMT & \textbf{S}urface-\textbf{m}ount \textbf{t}echnology \\
		DC & \textbf{D}irect \textbf{c}urrent \\
		PU & \textbf{P}oly(\textbf{u}rethane) \\
		PVC & \textbf{P}oly(\textbf{v}inyl \textbf{c}hloride) \\
		DI & \textbf{D}e\textbf{i}onized (water) \\
		GNU GPL & \textbf{GNU} \textbf{G}eneral \textbf{P}ublic \textbf{L}icense \\
		CERN OHL & \textbf{CERN} \textbf{O}pen \textbf{H}ardware \textbf{L}icense \\
		CC & \textbf{C}reative \textbf{C}ommons \\
		\bottomrule
	\end{tabulary}
\end{center}

\clearpage
\twocolumn
\printbibliography
\end{document}